\documentclass[aps,superscriptaddress,floatfix,nobibnote]{revtex4}

\usepackage{subfigure}
\usepackage{epsfig}
\usepackage{epstopdf}
\usepackage{color}
\usepackage{graphicx}
\usepackage{longtable}
\usepackage{CJK}
\usepackage{amsmath,bbold}
\usepackage{float}
\usepackage{physics}

\ifx\pdfoutput\undefined
\usepackage[dvips]{graphicx}
\usepackage[dvipdfm,
  pdfstartview=FitH,
     CJKbookmarks=true,
          bookmarksnumbered=true,
          bookmarksopen=true,
         colorlinks=true,
          pdfborder=002,
          citecolor=blue%
           ]{hyperref}
\AtBeginDvi{} 
\else
\usepackage[
          pdfstartview=FitH,
          CJKbookmarks=true,
          bookmarksnumbered=true,
           bookmarksopen=ture,
           colorlinks=true,
           citecolor=blue
           ]{hyperref}
\begin{document}

\title{New Procedure for Evaluation of U(3) Coupling and Recoupling Coefficients}

\author{Phong Dang}
\email[Corresponding author: ]{pdang5@lsu.edu}
\homepage[]{https://www.researchgate.net/profile/Phong-Dang-3}
\affiliation{Department of Physics and Astronomy, Louisiana State University, Baton Rouge, LA 70803-4001, USA}

\author{Jerry P. Draayer}
\email[]{draayer@lsu.edu}
\homepage[]{https://www.researchgate.net/profile/J-Draayer}
\affiliation{Department of Physics and Astronomy, Louisiana State University, Baton Rouge, LA 70803-4001, USA}

\author{Feng Pan}
\email[]{daipan@dlut.edu.cn}
\homepage[]{https://www.researchgate.net/profile/Feng-Pan}
\affiliation{Department of Physics, Liaoning Normal University, Dalian 116029, China}
\affiliation{Department of Physics and Astronomy, Louisiana State University, Baton Rouge, LA 70803-4001,
USA}

\author{Kevin S. Becker}
\email[]{kbeck13@lsu.edu}
\affiliation{Department of Physics and Astronomy, Louisiana State University, Baton Rouge, LA 70803-4001, USA}

\begin{abstract}

\vskip .2cm


{\bf Abstract}: A simple method to calculate Wigner coupling coefficients and Racah recoupling coefficients for U(3) in two group-subgroup chains is presented. While the canonical $\rm U(3)\supset U(2)\supset U(1)$ coupling and recoupling coefficients are applicable to any system that respects U(3) symmetry, the $\rm U(3) \supset SO(3)$ coupling coefficients are more specific to nuclear structure studies. This new procedure precludes the use of binomial coefficients and alternating sums which were used in the 1973 formulation of Draayer and Akiyama, and hence provides faster and more accurate output of requested results. The resolution of the outer multiplicity is based on the null space concept of the U(3) generators proposed by Arne Alex et al., whereas the inner multiplicity in the angular momentum subgroup chain is obtained from the dimension of the null space of the SO(3) raising operator. A C++ library built on this new methodology will be published in a complementary journal that specializes in the management and distribution of such programs.
\end{abstract}

\maketitle

\section{INTRODUCTION}

Mathematical topics underpinning what is commonly referred to as group theory emerged and evolved from set theory, the origin of which can be traced back to an early (1874) work by Georg Cantor, ``On a Property of the Collection of All Real Algebraic Numbers'' \cite{Cantor1874}. See also ``A History of Set Theory" \cite{Johnson1972history} and a more extended historical perspective given by Joseph Dauben \cite{Dauben1979}, who lauded Cantor for his ground-breaking work in the set theory arena. In Dauben's book, he suggested that set theory was initially used to introduce mathematics students to the triplet of the logical operations (NOT, AND, OR), which is of more modern interest because Boolean logic of this type was ultimately used -- nearly a century later -- to underpin the earliest development of computer programming languages. In modern times, group theory as understood and practiced might best be described as a study of relationships between and among sets of intertwined features, see \cite{Kolmogorov1970}.

\vskip.2cm

Jumping forward to the simplest, best known and most studied of all groups are the continuous unitary groups, U(N), where N specifies the number of degrees of freedom of physical systems of interest; hence U(1) is simply a linear 1-dimensional model space, U(2) is the well-known 2-dimensional model space which is commonly used to gain an algebraic description of simple harmonic motion of the pendulum type, and U(3) is used to describe a system with three degrees of freedom, etc. The most common usage of U(3) can be found in subatomic (nuclear and particle) physics. An early application of the former can be found in the pioneering work of Eugene Wigner, who in 1937 \cite{Wigner1937PR} proposed what he christened a ``Supermultiplet Symmetry" for the study of the structure of atomic nuclei, which at that time was clearly a theory ahead of its time. Nonetheless, it was a theory for which Wigner was ultimately awarded a Nobel Prize in Physics in 1963, along with nuclear physicists Maria Goeppert Mayer and J. Hans D. Jensen for their efforts in the mid-1940s that were focused on advancing the so-called single-particle shell model. What Wigner advanced is now usually referred to as Wigner's Supermultiplet Theory, wherein he proffered his famous factorization lemma in which the spatial and spin-isospin parts of a nuclear system are separated into their respective U(3) and U(4) parts. Also important to this cryptic background story is the work of J. P. Elliott, who in 1958 was the first to show the importance of U(3) group within the context of advancing a nuclear many-particle theory that was underpinned by the U(3) symmetry \cite{Elliott1958PRSA,Elliott1958PRSA-II,Elliott1963}. 

\vskip.2cm

Many researchers have contributed to developments in this arena, which requires knowledge of the coupling as well as recoupling of representations of U(3) and U(4) respectively, that are necessary for microscopic analyses of the structure of atomic nuclei.  Indeed, in 1973, Akiyama and Draayer published a pair of papers, the first of which was theoretically oriented and a second complementary one that proffered an algebraic code for calculating U(3) coupling and recoupling coefficients \cite{Draayer1973JMP,Akiyama1973} which has to date stood the test of time with various improved versions using extended precision \cite{Rowe2000JMP,Bahri2004CPC} and a modern programming language \cite{Dytrych2021CPC}. However, with the advent of modern computational resources in the mid-1990s, along with follow-on algebraic algorithmic developments, the 1973 package can now be updated to sidestep some of the limiting sticky wickets issues in favor of an improved and streamlined methodology, one that should enable moving upward towards the provisioning of a shell-model description of heavier and more complex atomic nuclei.  Since all of the U(N) groups are ubiquitous stand-alone structures that are divorced of any specific physical application of the theory, what follows is driven by the implementation of modern methodologies for determining the coupling and recoupling coefficients of U(3), which itself anticipates a soon-to-follow paper focused on U(4), that is necessary for a comprehensive utilization of Wigner's Supermultiplet Symmetry; one that should serve to respect and honor Wigner's contribution to science at a double-jubilee celebration of Wigner's efforts. Those close to Wigner knew that he himself was known to poetically proffer on numerous occasions that physics is really about finding ``Simplicity within Complexity''! 

\vskip.2cm

In this article, Section II is dedicated to a review of the basic features that define the unitary group U(3) and its irreducible representations. This is followed by Section III which is used to establish the labelling conventions for the canonical group-subgroup chain of U(3); more specifically, the labelling of basis states and how they are related via the actions of the generators of U(3). Section IV provides for a procedure for calculating coupling coefficients in the canonical scheme which was derived specifically for U(3) from a more general methodology for SU(N) advanced by Arne Alex, et al. \cite{Alex2011JMP}. Section V focuses on determining the recoupling coefficients for U(3); that is, the U(3) equivalents of the much studied 6-$j$ and 9-$j$ coefficients for SO(3), which we will call 6-U(3) and 9-U(3) coefficients to clearly distinguish them from the recoupling coefficients of SO(3). And Section VI is a review of the methodology used for evaluating Wigner $\rm SU(3)\supset SO(3)$ coefficients, which is very different from the 1973 methodology \cite{Draayer1973JMP}. And lastly, in the concluding Section VII a summary of our methodologies is given along with a brief snapshot of potential follow-on usages of the various methodologies introduced in this paper.

\vskip.2cm

\section{The Unitary Group U(3) and its Irreducible Representations}

The unitary group U(3) is a rank-2 group with an algebra that is spanned by nine generators: $E_{ij}$ (where $i,j=1,2,3$) which satisfy the following commutation relations and conjugation property:
\begin{align}
    [E_{ij},E_{jk}] = E_{ik} \text{ if } i \ne k \text{;  } [E_{ij},E_{ji}] = E_{ii} - E_{jj};   [E_{i_1 j_1},E_{i_2 j_2}] = \text{ 0 if } j_1 \ne i_2, i_1 \ne j_2 \text{; and } E_{ij}^\dagger &= E_{ji}.
    \label{eq:properties}
\end{align}
By taking various combinations of these generators, one can identify different subgroups of U(3), and hence group-subgroup chains that can be used to describe different physical phenomena. Nonetheless, it is important to mention the simplest group-subgroup chain up front:
\begin{align}
    \rm U(3) = U(1) \otimes SU(3).
\end{align}
The subgroup SU(3) of U(3) is a special unitary group composed of eight generators that remain after removing the first order Casimir invariant of U(3),  $C_1[{\rm U(3)}] = E_{11} + E_{22} + E_{33}$, which is the trace of a three-dimensional matrix representation of the nine generators of the U(3) group. 
\vskip.2cm
\begin{figure}[h!]
    \centering
    \includegraphics[scale=0.4]{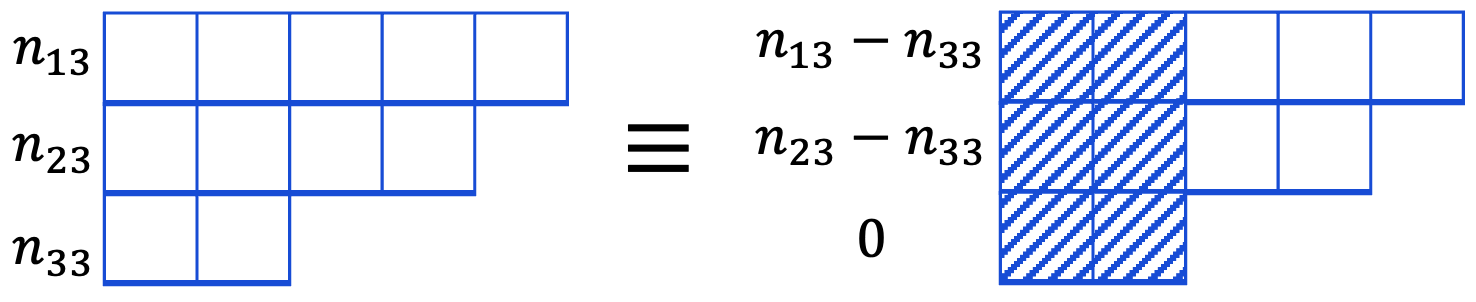}
    \caption{Young tableau representation of U(3) [on the lelft] and SU(3) irreps [on the right].}
    \label{fig:Young-tab}
\end{figure}

An irreducible representation of U(3), hereafter referred to as an irrep of U(3), is labeled by three quantum numbers, $[n_{13},n_{23},n_{33}]$, which are non-negative integers in decreasing order (i.e., $n_{13}\ge n_{23} \ge n_{33}$) and are often interpreted as the number of quanta along the three Cartesian axes in 3-space.  It is well known that two U(3) irreps $[n_{13},n_{23},n_{33}]$ and $[n_{13}+m,n_{23}+m,n_{33}+m]$ are basically equivalent \cite{Chen1989,Pan2016NPA}, hence an SU(3) irrep can be labeled by two quantum numbers $[n_{13}'=n_{13}-n_{33},n_{23}'=n_{23}-n_{33}]$, where $[n_{13},n_{23},n_{33}]$ is a corresponding U(3) irrep. (In the work of Alex et al. \cite{Alex2011JMP}, this is referred to as ``normalizing" an irrep.) The irreps of U(3) and SU(3) can also be represented intuitively by Young tableaux, which contain three rows for U(3) and two rows for SU(3), and the number of boxes on each row is equal to the respective partition quantum number in the irrep, see FIG. \ref{fig:Young-tab}. In this representation, the reduction of a U(3) irrep to an SU(3) irrep is achieved by simply removing the completed three-deep columns appearing on the left hand side of the tableau.  From this one can see that when reducing from U(3) to SU(3), one loses ``information" about the total number of boxes that are being distributed among the three Cartesian axes; nonetheless, the relative distribution between them is still known via the $[n_{13}',n_{23}']$ labeling of the SU(3) irrep. However, with the value of the first-order Casimir invariant of U(3), one can restore the full accounting of the total number of boxes from the labelling of SU(3).

\section{The Canonical Subgroup Chain U(3)$\supset$U(2)$\supset$U(1)}

\begin{figure}[h!]
    \centering
\includegraphics[scale=0.6]{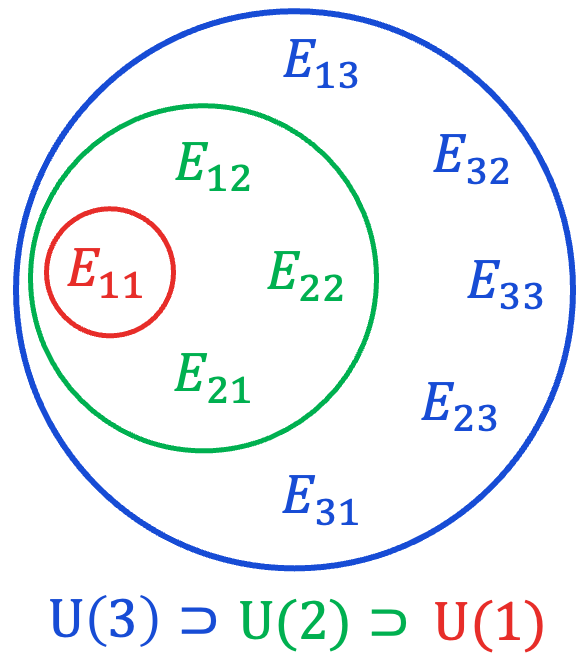}
    \caption{The canonical group-subgroup chains of U(3).}
    \label{fig:canonical_chain}
\end{figure}

Among the nine generators, $E_{ij}$ ($i,j=1,2,3$), of the U(3) group, there are four that generate all infinitesimal transformations of the unitary group U(2), namely $E_{ij}$ ($i,j=1,2$), out of which $E_{11}$ is exactly the only generator of the group U(1). Therefore, the natural reduction of the group U(3) is $\rm U(3) \supset U(2) \supset U(1)$, which is often referred to as a canonical reduction and has been studied thoroughly, e.g., see \cite{Louck1970JMP,Biedenharn1972JMP1,Biedenharn1972JMP2,Louck1973JMP,Alex2011JMP}. A graphic illustration of this group-subgroup chain is given in FIG. \ref{fig:canonical_chain}.
 
In this scheme, the basis states are labelled by six quantum numbers organized in an elegant pattern known as \textit{Gel'fand-Tsetlin pattern} \cite{Gelfand1950} -- hereafter called Gelfand state -- which is written in three rows as follows  
\begin{equation*}
\ket{G} = \ket{\begin{array}{c}
n_{13},n_{23},n_{33}\cr
n_{12},n_{22}\cr
n_{11}
\end{array}}, \text{ where the \textit{betweenness} condition must be satisfied: } n_{ij} \ge n_{i,j-1} \ge n_{i+1,j}.
\end{equation*}
With the betweenness condition, one can write down different Gelfand states under the same irrep, which make up the so-called carrier space of the irrep. The total number of valid patterns that one can build from a given irrep is known as the dimension of the irrep (or its carrier space), which can be computed via the following formula:
\begin{align}
    \dim([n_{13},n_{23},n_{33}]) = (1+n_{13}-n_{23})(1+n_{23}-n_{33})\left(1+ \frac{n_{13}-n_{33}}{2} \right).
    \label{eq:dim1}
\end{align}
\vskip.2cm

For the SU(3) case, as described above, one simply needs to remove the full columns of the Young tableau representing the irrep, which means there is an exact correspondence under which the matrix representations associated with a given U(3) irrep and its corresponding SU(3) irrep are the same:
\begin{align}
    \ket{\begin{array}{c}
n_{13},n_{23},n_{33}\cr
n_{12},n_{22}\cr
n_{11}
\end{array}} = \ket{\begin{array}{c}
n_{13}-n_{33},n_{23}-n_{33},0\cr
n_{12}-n_{33},n_{22}-n_{33}\cr
n_{11}-n_{33}
\end{array}}.
\end{align}
\vskip.2cm

Although all the generators work within a single irrep and do not mix irreps together, their action on the basis states varies. To be specific, three generators when acting upon the basis state do not change the state itself, and simply read out the number of quanta distributed in each direction:
\begin{align}
    E_{11}\ket{G} &= n_{11} \ket{G} \nonumber\\
    E_{22}\ket{G} &= (n_{12} + n_{22} - n_{11}) \ket{G}     \label{eq:remain}\\
    E_{33}\ket{G} &= (n_{13}+n_{23}+n_{33}-n_{12}-n_{22}) \ket{G} \nonumber;
\end{align}
in other words, all the Gelfand states are eigenstates of the generators $E_{ii}$ ($i=1,2,3$), and the eigenvalues make up a so-called pattern weight (or p-weight for short) \cite{Alex2011JMP}: $[w_3,w_2,w_1]:=[n_{13}+n_{23}+n_{33}-n_{12}-n_{22},n_{12}+n_{22}-n_{11},n_{11}]$. It can be seen that a p-weight does not uniquely define a Gelfand state -- indeed, there can be multiple Gelfand states associated with the same p-weight. Furthermore, since the irrep is known, i.e., $n_{13},n_{23},n_{33}$ are given in the irrep, one can use an alternative weight with only two numbers, $[z_2,z_1]:=[n_{12}+n_{22},n_{11}]$, which is called the z-weight of a Gelfand state \cite{Alex2011JMP}. 

\vskip.2cm

\begin{figure}[h!]
    \centering
    \includegraphics[scale=0.5]{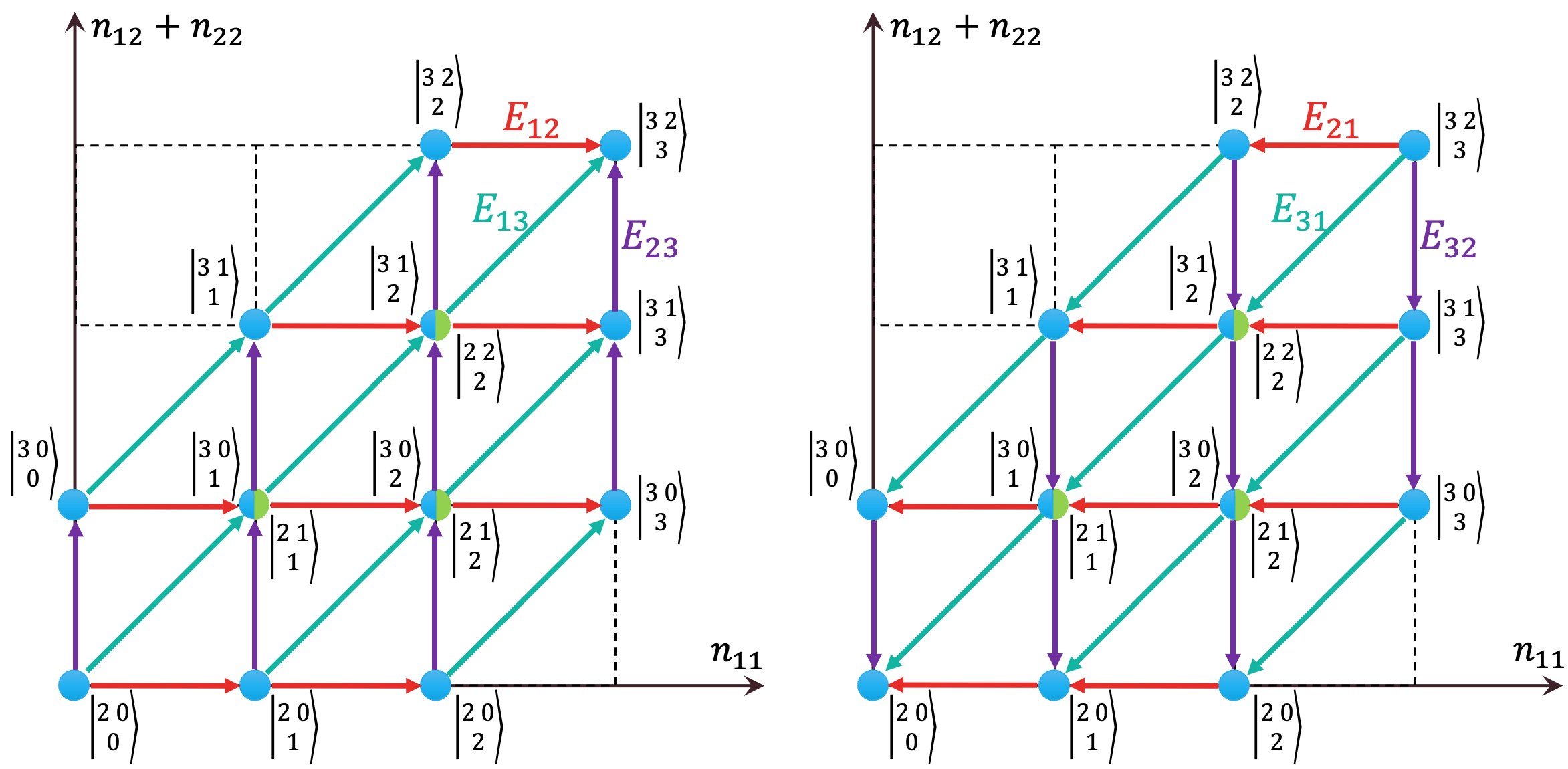}
    \caption{The z-weight diagram for the U(3) irrep $[3,2,0]$, with a demonstration of the action on all of the Gelfand states of the U(3) raising generators (left-hand panel) pushing upward towards the so-called highest weight state (sometimes referred to as the extremal state), and the action of lowering generators (right-hand panel), pushing downward towards the so-called lowest weight state (sometimes referred to as the conjugate of the highest-weight state), using colored arrows. Since the irrep is specified, the top row of the Gelfand patterns are not shown.}
    \label{fig:generators}
\end{figure}

Meanwhile, the other generators serve as \textit{raising} and \textit{lowering} operators that act on an individual entry of the state and change the state to one or two different states of the same irrep: 
\begin{align}
    E_{12}\ket{G} =& \sqrt{(n_{12}-n_{11})(n_{11}-n_{22}+1)} \ket{G + I_{11}}, \nonumber \\
    E_{21}\ket{G} =& \sqrt{(n_{12}-n_{11}+1)(n_{11}-n_{22})} \ket{G - I_{11}}, \nonumber \\
    E_{23}\ket{G} =& \sqrt{\frac{(n_{13}-n_{12})(n_{12}-n_{23}+1)(n_{12}-n_{33}+2)(n_{12}-n_{11}+1)}{(n_{12}-n_{22}+2)(n_{12}-n_{22}+1)}} \ket{G+I_{12}} \nonumber \\
            &+ \sqrt{\frac{(n_{13}-n_{22}+1)(n_{23}-n_{22})(n_{22}-n_{33}+1)(n_{11}-n_{22})}{(n_{12}-n_{22}+1)(n_{12}-n_{22})}} \ket{G+I_{22}},  \nonumber \\
    E_{32}\ket{G} =& \sqrt{\frac{(n_{13}-n_{12}+1)(n_{12}-n_{23})(n_{12}-n_{33}+1)(n_{12}-n_{11})}{(n_{12}-n_{22}+1)(n_{12}-n_{22})}} \ket{G-I_{12}} \nonumber \\
                &+ \sqrt{\frac{(n_{13}-n_{22}+2)(n_{23}-n_{22}+1)(n_{22}-n_{33})(n_{11}-n_{22}+1)}{(n_{12}-n_{22}+2)(n_{12}-n_{22}+1)}} \ket{G-I_{22}}, \label{eq:raise-lower}\\
    E_{13}\ket{G} =& \sqrt{\frac{(n_{13}-n_{12})(n_{12}-n_{23}+1)(n_{12}-n_{33}+2)(n_{11}-n_{22}+1)}{(n_{12}-n_{22}+2)(n_{12}-n_{22}+1)}} \ket{G+I_{12}+I_{11}} \nonumber\\
                &- \sqrt{\frac{(n_{13}-n_{22}+1)(n_{23}-n_{22})(n_{22}-n_{33}+1)(n_{12}-n_{11})}{(n_{12}-n_{22}+1)(n_{12}-n_{22})}} \ket{G+I_{22}+I_{11}}, \nonumber\\
    E_{31}\ket{G} =& \sqrt{\frac{(n_{13}-n_{12}+1)(n_{12}-n_{23})(n_{12}-n_{33}+1)(n_{11}-n_{22})}{(n_{12}-n_{22}+1)(n_{12}-n_{22})}} \ket{G-I_{12}-I_{11}} \nonumber\\
                &- \sqrt{\frac{(n_{13}-n_{22}+2)(n_{23}-n_{22}+1)(n_{22}-n_{33})(n_{12}-n_{11}+1)}{(n_{12}-n_{22}+2)(n_{12}-n_{22}+1)}} \ket{G-I_{22}-I_{11}}, \nonumber
    \label{eq:raising-lowering}
\end{align}
where we have introduced a notation for convenience $\ket{G \pm I_{ij}}$, which basically means that the entry $n_{ij}$ of the Gelfand state $\ket{G}$ that is being acted upon is modified by 1, the plus sign corresponds to a raise, whereas the minus sign displays a decrease.  Again, this shorthand notation implies neither addition nor subtraction between the Gelfand states and should not be understood in that way. Let us introduce further shorthand notations to these equations
\begin{align}
    E_{12}\ket{G} =& e_{12}(G) \ket{G + I_{11}}, \nonumber\\
    E_{21}\ket{G} =& e_{21}(G) \ket{G - I_{11}}, \nonumber\\
    E_{23}\ket{G} =& e_{23,1}(G) \ket{G + I_{12}} + e_{23,2}(G) \ket{G + I_{22}}, \\
    E_{32}\ket{G} =& e_{32,1}(G) \ket{G - I_{12}} + e_{32,2}(G) \ket{G - I_{22}}, \nonumber\\
    E_{13}\ket{G} =& e_{13,1}(G) \ket{G + I_{12} + I_{11}} + e_{13,2}(G) \ket{G + I_{22} + I_{11}}, \nonumber\\
    E_{31}\ket{G} =& e_{31,1}(G) \ket{G - I_{12} - I_{11}} + e_{31,2}(G) \ket{G - I_{22} - I_{11}}, \nonumber
\end{align}
which will prove useful in follow-on sections. FIG. \ref{fig:generators} is a z-weight diagram that serves as a demonstration for the action of the raising and lowering generators on Gelfand basis states.

\section{U(3)$\supset$U(2)$\supset$U(1) Coupling Coefficients}

In exploring different physical phenomena, it often occurs that our physical system of interest is not a simple one, rather it is composite and made up of smaller constituents. Therefore, if the U(3) symmetry is valid for both individual subsystems and the entire system as a whole, one encounters the problem of coupling U(3) irreps that belong to each subsystem to obtain the irrep describing the whole system. 

\vskip.2cm

In the simplest case, one needs to couple two U(3) irreps, e.g., binary clustering of atomic nuclei \cite{Darai2012PRC,Dang2023PRC}, each of which has an associated carrier space. The coupling results in various irreps, each has its own carrier space; therefore, it is vital to know the transformation from the uncoupled spaces to the coupled one. In the SO(3) case, the transformation is given via the Clebsch-Gordan coefficients (abbreviated as CGC) $\braket{j_1m_1,j_2m_2}{JM}$, or its equivalent 3-$j$ symbols introduced by Wigner. Similarly, when one works with the U(3) group, the transformation is also given by CGC's, which we call Wigner 3-U(3) coefficients, and can be understood as the expansion coefficients of a Gelfand state in the coupled carrier space in terms of those in the uncoupled carrier spaces as follows,
\begin{align}
    \ket{G''}_\rho = \sum_{G,G'} \braket{G;G'}{G''}_\rho \ket{G}\ket{G'} = \sum_{G,G'} C_{G,G'}^{G'',\rho} \ket{G}\ket{G'},
    \label{eq:cano-expansion}
\end{align}
where $C_{G,G'}^{G'',\rho}$ are the canonical $\rm U(3)\supset U(2) \supset U(1)$ CGC's. The integer $\rho$ is simply a counter for multiple occurrences of $\ket{G''}$; the details of how to find the maximal value of $\rho$ will be given below. Note that the procedure in this section is similar to a more general procedure for U(N) proposed by Alex et al. \cite{Alex2011JMP}, however, explicit formulation is derived specifically for the U(3) case. 

\subsection{Direct Product of Two U(3) Irreps}
The direct product of two U(3) irreps, $[n_{13},n_{23},n_{33}]$ and $[n_{13}',n_{23}',n_{33}']$, in general can be decomposed into a direct sum of U(3) irreps, each of which can have multiple occurrences:
\begin{align}
    [n_{13},n_{23},n_{33}] \otimes [n_{13}',n_{23}',n_{33}'] = \oplus \rho_{\max} [n_{13}'',n_{23}'',n_{33}''], 
\end{align}
where the outer multiplicity $\rho_{\max}$ is the number of times that the coupled irrep $[n_{13}'',n_{23}'',n_{33}'']$ occurs in the decomposition. 

\vskip.2cm

A detailed algorithm to obtain both the coupled irrep and the outer multiplicity thereof based on the Littlewood-Richardson rule can be found in section VIII of \cite{Alex2011JMP} together with an explicit example, $[2,1,0]\otimes [2,1,0]$, which can be found in Table I therein. It is not the focus of this article to present this well-known method.  

\vskip.2cm

As noted above, the generators $E_{ij}$ only work within the associated carrier space of the same irrep. When coupling two irreps together, it is possible to express the generators in the coupled carrier space in terms of those from the individual carrier spaces as follows
\begin{align}
    E_{ij}'' = E_{ij}\otimes \mathbf{1}' + \mathbf{1}\otimes E_{ij}',
\end{align}
where $\mathbf{1}$ is the identity of the corresponding carrier space. One can double check easily that when the generators in the coupled carrier space are built in this way, they satisfy all of the properties given in Eq. (\ref{eq:properties}).

\subsection{Selection Rule for U(3)$\supset$U(2)$\supset$U(1) CGC's}
Applying the generator $E_{33}'' = E_{33}\otimes\mathbf{1}' + \mathbf{1}\otimes E_{33}'$ on  both sides of Eq. (\ref{eq:cano-expansion}), we obtain
\begin{align}
    w_3'' \ket{G''}_\rho = \sum_{G,G'} C_{G,G'}^{G'',\rho} (w_3 + w_3')\ket{G}\ket{G'},
\end{align}
which implies that $w_3'' = w_3 + w_3'$. Similarly for $E_{11}$ and $E_{22}$, we have $w_2''=w_2+w_2'$ and $w_1''=w_1+w_1'$. (The $[w_3,w_2,w_1]$ triplet is the p-weight of a Gelfand state.) Hence, these equations serve not only to determine which $\ket{G}$ and $\ket{G'}$ are involved in the expansion, Eq. (\ref{eq:cano-expansion}), but also as a selection rule for the CGC's, $C_{G,G'}^{G'',\rho}$.  That is,
\begin{align}
    C_{G,G'}^{G'',\rho} = 0 \text{ when } w_i'' \ne w_i+w_i' \text{ with } i=1,2,3.
\end{align}
Note that the condition $w_3''=w_3+w_3'$ is redundant if the other two are already satisfied, and therefore does not need to be checked when expanding $\ket{G''}$ in terms of $\ket{G}$ and $\ket{G'}$, i.e., we are left with only two conditions
\begin{align}
    n_{11}'' &= n_{11} + n_{11}', \nonumber\\
    n_{12}'' + n_{22}'' &= n_{12} + n_{22} + n_{12}' + n_{22}'.
\end{align}

\subsection{U(3)$\supset$U(2)$\supset$U(1) CGC's for the Highest Weight States}

The highest weight state -- or the stretched state -- for a given U(3) irrep $[n_{13},n_{23},n_{33}]$ is defined as
\begin{align}
    \ket{HW} := \ket{\begin{array}{c}
n_{13},n_{23},n_{33}\cr
n_{13},n_{23}\cr
n_{13}
\end{array}},
\end{align}
which must vanish under the action of the three generators $E_{12}$, $E_{23}$, and $E_{13}$, see FIG. \ref{fig:generators}. In other words, the highest weight states belong to the null space of those three generators. However, owing to the commutation relation $[E_{12},E_{23}]=E_{13}$, any state that simultaneously belongs to the null space of $E_{12}$ and $E_{23}$ will automatically belong to the null space of $E_{13}$. 

\vskip.2cm

By applying $E_{12}''=E_{12}\otimes \mathbf{1}' + \mathbf{1}\otimes E_{12}'$ on both sides of Eq. (\ref{eq:cano-expansion}) with $\ket{G''}_\rho = \ket{HW''}_\rho$, we get the following linear equation
\begin{align}
    \sum_{G,G'} C_{G,G'}^{HW'',\rho} \left(e_{12}(G) \ket{G+I_{11}}\ket{G'} + e_{12}(G') \ket{G}\ket{G'+I_{11}} \right) = 0.
    \label{eq:e12}
\end{align}
Similarly for $E_{23}''=E_{23}\otimes \mathbf{1}' + \mathbf{1}\otimes E_{23}'$, we have
\begin{multline}
    \sum_{G,G'} C_{G,G'}^{HW'',\rho} \biggl( e_{23,1}(G) \ket{G+I_{12}}\ket{G'} 
            + e_{23,2}(G) \ket{G+I_{22}}\ket{G'} \\
            + e_{23,1}(G') \ket{G}\ket{G'+I_{12}} + e_{23,2}(G') \ket{G}\ket{G'+I_{22}} \biggl) =0.
            \label{eq:e23}
\end{multline}
It can be realized that the action of the generators on different Gelfand states can result in the same Gelfand state, but with different coefficients. By grouping the coefficients in front of the same state which are resulted from the action of the raising generators, one can obtain a homogeneous linear equation whose unknowns are the CGC's of the highest weight state. This suggests that Eqs. (\ref{eq:e12}) and (\ref{eq:e23}) provide us a set of homogeneous linear equations, which can be written in a compact matrix form
\begin{align}
    \mathbf{P}(HW'')\mathbf{C}^\rho = \mathbf{0},
\end{align}
where the matrix elements of $\mathbf{P}(HW'')$ are given in Eqs. (\ref{eq:e12}) and (\ref{eq:e23}), and all the CGC's $C_{G,G'}^{HW'',\rho}$ are entries of the column vector $\mathbf{C}^\rho$  for all $\rho = 1,2,...,\rho_{\max}$. With modern techniques and algorithms, it is possible to obtain simultaneously all CGC's for all multiplicities $\rho$ by solving the null space of the matrix $\mathbf{P}(HW'')$, and the value of $\rho_{\max}$ is resolved by the dimensionality of the null space thereof. Nevertheless, it often happens that the solution of the null space of a matrix does not contain orthonormalized column vectors, thus a Gram-Schmidt procedure can be carried out to transform the result to a set of orthogonal column vectors of unit norm. Furthermore, in the presence of a single outer multiplicity, the coupling (thus the solution of the null space) is unique; on the contrary, when there are multiple occurrences of the coupled irrep, the expansion coefficients are not unique anymore and are somewhat arbitrary \cite{Alex2011JMP}, in which case, nonetheless, all solutions are still equivalent because one can be obtained from another via a unitary or orthogonal transformation. 

\vskip.2cm

This way of calculating CGC's via solving the null space also provides us with a very convenient way to fix the phase of the CGC's. To be more specific, we only need to set the top coefficient of each column vector solution of the null space to be positive, and the sign of all other coefficients can be determined by the relative phase with respect to the top one. Once the signs of all the stretched coefficients are chosen, the signs of all lower weight coupling coefficients simply follow the actions of the lowering generators.

\subsection{U(3)$\supset$U(2)$\supset$U(1) CGC's for the Lower Weight States}
It is natural to assume that from the highest weight state, one can go down using lowering generators $E_{21}''$ and $E_{32}''$. However, there is a challenge due to the fact that the action of the latter generators on any state $\ket{G''}_\rho$ (including $\ket{HW''}_\rho$) by and large returns two different Gelfand states, see Eqs. (\ref{eq:raise-lower}). Therefore, we will evaluate CGC's for all states of the same weight $[w_3'',w_2'',w_1'']$ at the same time. 

\vskip.2cm

Let $\ket{G''}_\rho$ be one of those states. A parent state of $\ket{G''}_\rho$ is a Gelfand state such that when it is acted upon by either $E_{21}''$ or $E_{32}''$, it returns back to $\ket{G''}_\rho$. There are three such states, $\ket{G''+I_{11}}_\rho$, $\ket{G''+I_{12}}_\rho$ and $\ket{G''+I_{22}}_\rho$, each of which can be expanded as follows,
\begin{align}
    \ket{G''+I_{11}}_\rho &= \sum_{G,G'} C_{G,G'}^{G''+I_{11},\rho} \ket{G}\ket{G'}, \nonumber\\
    \ket{G''+I_{12}}_\rho &= \sum_{G,G'} C_{G,G'}^{G''+I_{12},\rho} \ket{G}\ket{G'}, \\
    \ket{G''+I_{22}}_\rho &= \sum_{G,G'} C_{G,G'}^{G''+I_{22},\rho} \ket{G}\ket{G'},\nonumber
\end{align}
where the expansion coefficients are assumed to be known already. Then applying $E_{21}''=E_{21}\otimes \mathbf{1}' + \mathbf{1}\otimes E_{21}'$ on both sides of the first equation, and $E_{32}''=E_{32}\otimes \mathbf{1}' + \mathbf{1}\otimes E_{32}'$ on both sides of the other two equations, we arrive at
\begin{eqnarray}\nonumber
   & e_{21}(G''+I_{11})\ket{G''}_\rho = \\\nonumber
   &\sum_{G,G'}\, C_{G,G'}^{G''+I_{11},\rho} \bigg[e_{21}(G)\ket{G-I_{11}}\ket{G'} + e_{21}(G')\ket{G}\ket{G'-I_{11}} \bigg], \\\nonumber
   \\\nonumber
   &  e_{32,1}(G''+I_{12})\ket{G''}_\rho + e_{32,2}(G''+I_{12})\ket{G''+I_{12}-I_{22}}_\rho =\\\nonumber
   &\sum_{G,G'}\, C_{G,G'}^{G''+I_{12},\rho}\, \bigg[e_{32,1}(G)\ket{G-I_{12}}\ket{G'} + e_{32,2}(G)\ket{G-I_{22}}\ket{G'} \\
    &+ e_{32,1}(G')\ket{G}\ket{G'-I_{12}} + e_{32,2}(G')\ket{G}\ket{G'-I_{22}} \bigg],\\\nonumber
    \\\nonumber
    & e_{32,1}(G''+I_{22})\ket{G''+I_{22}-I_{12}}_\rho + e_{32,2}(G''+I_{22})\ket{G''}_\rho = \\\nonumber
    &\sum_{G,G'}\, C_{G,G'}^{G''+I_{22},\rho} \bigg[e_{32,1}(G)\ket{G-I_{12}}\ket{G'} + e_{32,2}(G)\ket{G-I_{22}}\ket{G'} \\\nonumber
    &+ e_{32,1}(G')\ket{G}\ket{G'-I_{12}} + e_{32,2}(G')\ket{G}\ket{G'-I_{22}} \bigg].
\end{eqnarray}
It can be seen that $\ket{G''}_\rho$, $\ket{G''+I_{12}-I_{22}}_\rho$, and $\ket{G''+I_{22}-I_{12}}_\rho$ are states of the same weight -- this is the motivation as to why we would like to compute CGC's for all states of the same weight simultaneously. On the grounds that we (presumably) already know the expansion of the parent states, and that the action of the lowering generators are given in Eqs. (\ref{eq:raise-lower}), all the coefficients $C_{G,G'}^{G'',\rho}$ can be obtained easily by organizing these equations into a matrix equation and then inverting the matrix on the left hand side which includes $e_{21}(G''+I_{11})$, $e_{32,1}(G''+I_{12})$,  $e_{32,2}(G''+I_{12})$, $e_{32,1}(G''+I_{22})$, and $e_{32,2}(G''+I_{22})$ as matrix elements. Note that, as it was pointed out in \cite{Alex2011JMP}, these equations are in general not linearly independent, i.e., the solution is overdetermined; therefore, it is necessary to remove those equations that are dependent on each other prior to performing the matrix inversion, so that the matrix on the left hand side is square and can be inverted.

\section{U(3) Racah Recoupling Coefficients}

The Racah recoupling coefficients arise when more than two irreps are coupled together, and they provide us with a transformation between various orders of coupling. For a detailed review of the recoupling coefficients, see the appendices of \cite{Escher1998JMP} and references therein. In this section, we simply show the procedure to evaluate them. For convenience, let us introduce a shorthand notation for U(3) irreps: $\Gamma \equiv [n_{13},n_{23},n_{33}]$.

\subsection{Coupling of Three U(3) Irreps: 6-U(3) Recoupling Coefficients} 
\begin{figure}[h!]
    \centering
    \includegraphics[scale=0.5]{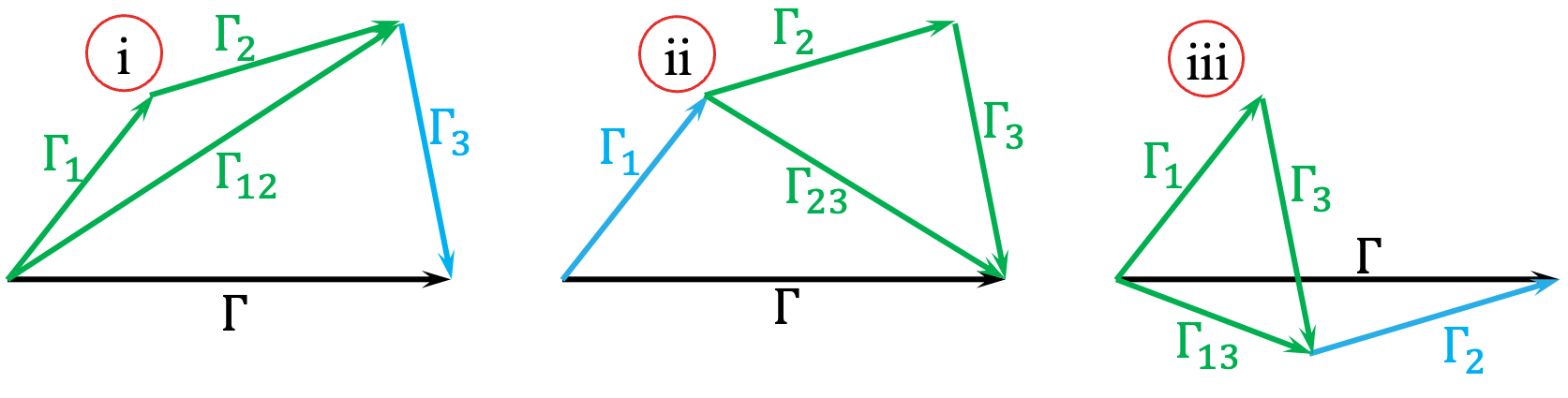}
    \caption{Three different orders of coupling three U(3) irreps  $\Gamma_1$, $\Gamma_2$ and $\Gamma_3$.}
    \label{fig:6-U(3)}
\end{figure}

When coupling three U(3) irreps $\Gamma_1$, $\Gamma_2$ and $\Gamma_3$ to a final irrep $\Gamma$, there are three different orders
\begin{align}
    (\Gamma_1 \otimes \Gamma_2) \otimes \Gamma_3, \text{  } \Gamma_1 \otimes (\Gamma_2 \otimes \Gamma_3), \text{  and  } (\Gamma_1 \otimes \Gamma_3) \otimes \Gamma_2,
\end{align}
as shown in FIG. \ref{fig:6-U(3)}. The transformation between scheme i) and scheme ii) is given by the so-called  $U$-coefficients, which can be obtain from the following relation
\begin{align}
    \sum_{\rho_{1,23}} C_{G_1,G_{23}}^{G,\rho_{1,23}} \times U(\Gamma_1\Gamma_2\Gamma\Gamma_3;\Gamma_{12}\rho_{12}\rho_{12,3},\Gamma_{23}\rho_{23}\rho_{1,23}) = \sum_{G_2 G_3 G_{12}} C_{G_1,G_2}^{G_{12},\rho_{12}} \times C_{G_{12},G_3}^{G,\rho_{12,3}} \times C_{G_2,G_3}^{G_{23},\rho_{23}}.
\end{align}
In contrast with \cite{Draayer1973JMP}, due to the way we compute CGC's that we present above, it is more computationally convenient to fix $\ket{G_{23}}=\ket{HW_{23}}$ and $\ket{G}=\ket{HW}$, while letting $\ket{G_1}$ run over its range;  then one can generate a set of linear equations with the $U$-coefficients as the unknowns. By solving this set of linear equations, all the $U$-coefficients for $\rho_{1,23}=1,2,...,\rho_{1,23}^{\max}$ can be found simultaneously. 

In an analogous manner, the so-called $Z$-coefficients provide the transformation between scheme i) and iii), which satisfy the relation
\begin{align}
    \sum_{\rho_{13,2}} C_{G_{13},G_2}^{G,\rho_{13,2}} \times Z(\Gamma_2\Gamma_1\Gamma\Gamma_3;\Gamma_{12}\rho_{12}\rho_{12,3},\Gamma_{13}\rho_{13}\rho_{13,2}) = \sum_{G_1 G_3 G_{12}} C_{G_1,G_2}^{G_{12},\rho_{12}} \times C_{G_{12},G_3}^{G,\rho_{12,3}} \times C_{G_1,G_3}^{G_{13},\rho_{13}}.
    \label{eq:Z6}
\end{align}
As suggested in \cite{Millener1978JMP}, by fixing $\ket{G_{13}}=\ket{HW_{13}}$ and $\ket{G}=\ket{HW}$ while letting $\ket{G_2}$ run over its range, one can solve a set of linear equations to obtain simultaneously the $Z$-coefficients for all $\rho_{13,2}=1,2,...,\rho_{13,2}^{\max}$.

\subsection{Coupling of Four U(3) Irreps: 9-U(3) Recoupling Coefficients}

When coupling four U(3) irreps $\Gamma_1$, $\Gamma_2$, $\Gamma_3$ and $\Gamma_4$ to a final irrep $\Gamma$, there are more orders of coupling. The transformation between two schemes, $(\Gamma_1 \otimes \Gamma_2) \otimes (\Gamma_3 \otimes \Gamma_4)$ and $(\Gamma_1 \otimes \Gamma_3) \otimes (\Gamma_2 \otimes \Gamma_4)$, is provided via the so-called 9-U(3) recoupling coefficients, which can be evaluated from the 6-$U$- and 6-$Z$-coefficients with the following formula:
\begin{multline}
    \Biggl\{ \begin{array}{cccc}
         \Gamma_1& \Gamma_2& \Gamma_{12}& \rho_{12} \\
         \Gamma_3& \Gamma_4& \Gamma_{34}& \rho_{34} \\
         \Gamma_{13}& \Gamma_{24}& \Gamma& \rho_{13,24} \\
         \rho_{13}& \rho_{24}& \rho_{12,34}
    \end{array} \Biggl\} = \sum_{\Gamma_0, \rho_{13,2} \rho_{04} \rho_{12,3}} U(\Gamma_{13}\Gamma_2\Gamma\Gamma_4;\Gamma_{0}\rho_{13,2}\rho_{04},\Gamma_{24}\rho_{24}\rho_{13,24})  \times \\ Z(\Gamma_2\Gamma_1\Gamma_0\Gamma_3;\Gamma_{12}\rho_{12}\rho_{12,3},\Gamma_{13}\rho_{13}\rho_{13,2})
    \times U(\Gamma_{12}\Gamma_3\Gamma\Gamma_4;\Gamma_{0}\rho_{12,3}\rho_{04},\Gamma_{34}\rho_{34}\rho_{12,34}),
\end{multline}
where the $\Gamma_0$'s are obtained from coupling three U(3) irreps $\Gamma_1$, $\Gamma_2$ and $\Gamma_3$, in both schemes: $(\Gamma_1 \otimes \Gamma_2) \otimes \Gamma_3$ and $(\Gamma_1 \otimes \Gamma_3) \otimes \Gamma_2$.

\section{Application in Nuclear Physics -- SU(3)$\supset$SO(3)$\supset$SO(2) Wigner Coefficients}

\subsection{Physical Reduction in Nuclear Physics}

\begin{figure}[h!]
    \centering
    \includegraphics[scale=0.3]{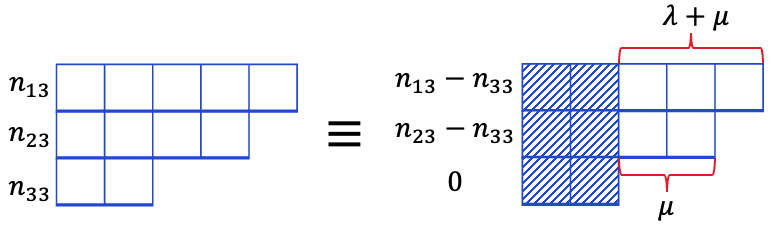}
    \caption{Young tableau representation of Elliott SU(3) labels.}
    \label{fig:elliott-su3}
\end{figure}

A more familiar representation of the special unitary group SU(3) in nuclear physics is given by two quantum numbers
\begin{align}
    \lambda = n_{13}'-n_{23}' = n_{13}-n_{23} \text{  and  } \mu = n_{23}' = n_{23} - n_{33},
\end{align}
which was introduced by Elliott in 1958 \cite{Elliott1958PRSA,Elliott1958PRSA-II} to describe collective rotational features of atomic nuclei; the Young tableau representation of a given $(\lambda,\mu)$ is demonstrated in FIG. \ref{fig:elliott-su3}, wherein the  $(\lambda,\mu)$ pairs refer to a counting of overhanging boxes.  Despite removing the first-order Casimir invariant, the dimension of a U(3) irrep and its SU(3) counterpart remains unchanged, and is provided via the following formula \cite{Draayer1968NPA}
\begin{align}
    \dim(\lambda,\mu) = (1+\lambda)(1+\mu)\left(1+ \frac{\lambda+\mu}{2} \right).
    \label{eq:dim2}
\end{align}

\begin{figure}[h!]
    \centering
    \includegraphics[scale=0.6]{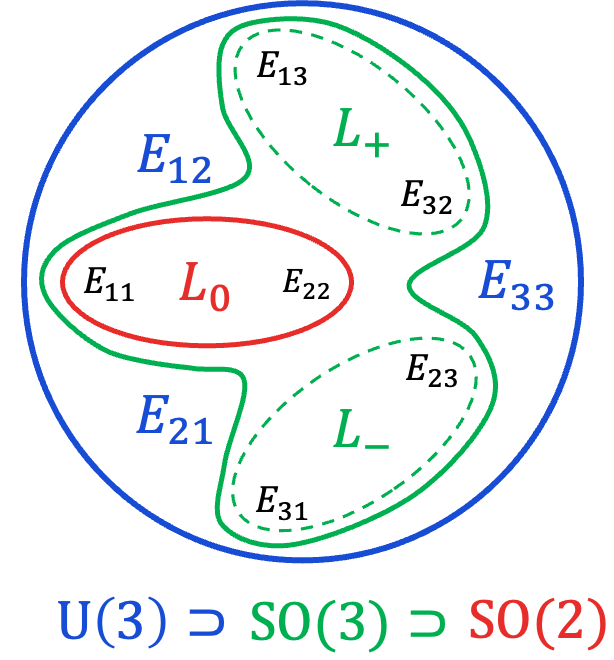}
    \caption{The group-subgroup chain realization of U(3) in nuclear physics.}
    \label{fig:phys-chain}
\end{figure}

In Elliott's model, it is the ``physical" group-subgroup chain $\rm SU(3)\supset SO(3) \supset SO(2)$ that is important, and the generators are constructed in the following way:
\begin{align}
    L_+ = \sqrt{2}(E_{13}+E_{32}) \text{, } L_0 = E_{11}-E_{22} \text{, } L_- = \sqrt{2}(E_{31}+E_{23}),
\end{align}
which are angular momentum operators that generate all infinitesimal transformations belonging to the groups SO(3) and SO(2). A graphic illustration of the physical reduction can be found in FIG. \ref{fig:phys-chain}. The action of the generators on the basis states is very simple: $L_0$ reads out the third component $M$ of the angular momentum (which is often thought of as the $z$-component in 3-space), $L_\pm$ are raising and lowering operators that increases and decreases the third component $M$ by 1, respectively. Since the group SU(3) is a group of rank 2, whereas the group SO(3) is a rank-1 group, the reduction from SU(3) down to SO(3) may bring about multiple occurrences of the angular momentum $L$, which Elliott resolved by employing a third label $K$, interpreted as the projection of the angular momentum $L$ on the symmetry axis of the nucleus itself. The reduction rule is as follows: for a given SU(3) irrep $(\lambda,\mu)$, 
\begin{align*}
    K = \min(\lambda,\mu), \min(\lambda,\mu)-2, ..., 0 \text{ or } 1;
\end{align*}
then for each $K$, there are two scenarios
\begin{align*}
    K>0 &\text{: } L = K,K+1,...,K+\max(\lambda,\mu), \\
    K=0 &\text{: } L = \max(\lambda,\mu),\max(\lambda,\mu)-2, ..., 0 \text{ or } 1.
\end{align*}

\vskip.2cm
However, in this paper, we employ a third label (similar to Draayer and Akiyama \cite{Draayer1973JMP}) denoted by $\kappa$, which simply serves as a counter for multiple occurrences of $L$ within the same U(3) irrep, i.e., the basis states are labeled by
\begin{align}
    \ket{\begin{array}{c}
n_{13},n_{23},0\cr
\kappa L M
\end{array}} \equiv \ket{(\lambda,\mu)\kappa L M} \text{ with } \lambda = n_{13}-n_{23}, \mu = n_{23}.
\end{align}
There is an explicit formula to determine the maximal value of $\kappa$, which is the so-called inner multiplicity of $L$ for a given SU(3) irrep $[n_{13},n_{23}]$ \cite{Pan2016NPA}: 
\begin{align}
    \kappa_{\max}(L) = \text{multi}([n_{13},n_{23}],L) = \text{IntP}\left[\frac{1}{2}(n_{13}-L) \right] - \text{IntP}\left[\frac{1}{2}(n_{13}-n_{23}+1-L) \right] - \text{IntP}\left[\frac{1}{2}(n_{23}+1-L) \right] +1,
\end{align}
where
\begin{align}
    \text{IntP}[x] = \begin{cases}
        \text{Int}[x] & \text{ if } x \ge 0\\
        0 & \text{ if } x<0
    \end{cases},
\end{align}
with $\text{Int}[x]$ returning the integer part of $x$. Note that the label $\kappa$ merely serves as a counter and does not carry any significant physical property like the label $K$ that characterizes rotational bands; nonetheless, should it be needed, one can include the operator $K^2$ via $X_3$ and $X_4$, which are the so-called Integrity Basis Operators of $\rm SU(3)$, see \cite{Rosensteel1977AnnP,Naqvi1990NPA} and Chapter 5 of \cite{Kota2020}.

\vskip.2cm

As we argued previously in Section IV, the atomic nucleus itself is a composite system with nucleons as constituents; therefore, there is a need to couple U(3) irreps and the basis states thereof. Owing to the independence of the recoupling coefficients from the subgroup chains, the formulas presented in Section V are still valid and do not need to be reviewed. However,  the Wigner $\rm SU(3) \supset SO(3)$ coupling coefficients are different; nonetheless, but we can take advantage of the canonical coupling coefficients from Section IV. In what follows, first we will discuss the transformation between the two sets of basis states -- \textit{physical} vs. \textit{canonical} -- followed by coupling coefficients in the physical basis via the transformation. Note that our labelling of basis states remain canonical, i.e., $[n_{13},n_{23},n_{33}]$ for U(3) and $[n_{13}'=n_{13}-n_{33},n_{23}'=n_{23}-n_{33}]$ for SU(3), in lieu of Elliott's labels $(\lambda,\mu)$.

\subsection{Transformation between the Physical and the Canonical Bases}

The construction of the $\rm SU(3) \supset SO(3)$ basis states from the canonical $\rm U(3) \supset U(2) \supset U(1)$ ones is presented in detail in \cite{Pan2016NPA}. In this subsection, we will briefly review the methodology presented therein that utilizes the null space concept and is drastically different from the hypergeometric function projection technique used in \cite{Draayer1973JMP}.

\vskip.2cm

As pointed out in \cite{Pan2016NPA}, there is a freedom for the starting point. One can either construct the highest weight state ($M=L$) or the lowest weight state ($M=-L$), and then generate all other states via applying consecutively the lowering or raising operators of SO(3). Here, we follow the standard procedure to evaluate SO(3) Clebsch-Gordan coefficients, that is, we start with the highest weight state
\begin{align*}
    \ket{\begin{array}{c}
n_{13},n_{23},0\cr
\kappa L M=L
\end{array}},
\end{align*}
and once it is known we apply the lowering operator to obtain other states
\begin{align}
    \ket{\begin{array}{c}
n_{13},n_{23},0\cr
\kappa L M
\end{array}} = \sqrt{\frac{(L+M)!}{(2L)!(L-M)!}} (L_-)^{L-M} \ket{\begin{array}{c}
n_{13},n_{23},0\cr
\kappa L M=L
\end{array}} \text{, where } L_- = \sqrt{2}(E_{31}+E_{23}).
\label{eq:lower-weight}
\end{align}

\vskip.2cm

For a given SU(3) irrep $[n_{13},n_{23}]$, the allowed values of $L$ are $0,1,...,n_{13}$. Then, for each $L$, the highest weight basis state can be expanded in the canonical basis as follows,
\begin{align}
    \ket{\begin{array}{c}
n_{13},n_{23},0\cr
\kappa L M=L
\end{array}} = \sum_{t=0}^{n_{23}} \sum_{q=q_{\min}(L,t)}^{q_{\max}(L,t)} C^{\kappa}_{qt}([n_{13},n_{23},0]LM=L) \ket{\begin{array}{c}
n_{13},n_{23},0\cr
L+2q-t,t \cr
L+q
\end{array}},
\label{eq:expansion}
\end{align}
where 
\begin{align}
    q_{\min}(L,t) = \max\left[t, \text{IntM}\left[\frac{1}{2}(t-L+n_{23}) \right] \right] \text{ and } q_{\max}(L,t) = \text{Int}\left[\frac{1}{2}(n_{13}-L+t) \right]
\end{align}
with 
\begin{align}
    \text{IntM}[x] = \begin{cases}
    \text{Int}[x] +1 & \text{ if } x> \text{Int}[x] \\
    \text{Int}[x]  & \text{ if } x= \text{Int}[x]
    \end{cases}.
\end{align}
The expansion coefficients $C^{\kappa}_{qt}([n_{13},n_{23},0]LM=L)$ can be found using the fact that the highest weight state, by definition, must vanish under the action of the raising operator $L_+$:
\begin{align}
    \frac{1}{\sqrt{2}}L_+ \ket{\begin{array}{c}
n_{13},n_{23},0\cr
\kappa L M=L
\end{array}} = (E_{13}+E_{32})\ket{\begin{array}{c}
n_{13},n_{23},0\cr
\kappa L M=L
\end{array}} = 0.   
\end{align}
Plugging into Eq. (\ref{eq:expansion}), we obtain
\begin{align}
 \sum_{t=0}^{n_{23}} \sum_{q=q_{\min}(L,t)}^{q_{\max}(L,t)} C^{\kappa}_{qt}([n_{13},n_{23},0]LM=L) (E_{13}+E_{32}) \ket{\begin{array}{c}
n_{13},n_{23},0\cr
L+2q-t,t \cr
L+q
\end{array}} = 0.
\label{eq:null-phys}
\end{align}
From subsection II.A, we know the action of the generators $E_{13}$ and $E_{32}$ on the canonical basis states; therefore, after carrying out the action of those generators, we obtain from Eq. (\ref{eq:null-phys}) a set of homogeneous linear equations with the expansion coefficients $C^{\kappa}_{qt}([n_{13},n_{23},0]LM=L)$ as the unknowns, which can be written in a compact matrix form
\begin{align}
    \mathbf{P}([n_{13},n_{23},0],L) \mathbf{C}^{\kappa} = \mathbf{0},
\end{align}
where we call $\mathbf{P}([n_{13},n_{23},0],L)$ the angular momentum projection matrix and the column vector $\mathbf{C}^{\kappa}$ contains all the expansion coefficients as its components. In this way, the problem of finding the transformation between the physical and canonical bases reduces to solving the null space of the angular momentum projection matrix. After solving the null space, all expansion coefficients associated with different $\kappa$'s are obtained simultaneously, and the inner multiplicity issue occurring in the reduction from SU(3) to SO(3) can be resolved by checking the dimension of the null space of $\mathbf{P}([n_{13},n_{23},0],L)$. It often happens that the solution of solving the null space does not contain orthonormalized column vectors, thus a Gram-Schmidt procedure can be carried out as we do for $\mathbf{P}(HW'')$ in Subsection IV.C. Note that if the inner multiplicity of $L$ is $\kappa_{\max}=1$, the solution is unique; however, when the same value of $L$ occurs multiple times, the expansion coefficients are not unique anymore and are somewhat arbitrary \cite{Pan1998JMP,Pan2014EPJP}, in which case, nonetheless, all solutions are still equivalent as pointed out before. Moreover, we can easily fix the phase of the expansion coefficients from the physical to the canonical basis by ensuring that the top elements of all the column vectors, $\mathbf{C}^\kappa$, are always positive, and then the signs of the other elements simply follow.

\vskip.2cm

In order to ensure that solving the null space of all the angular momentum projection matrices is carried out correctly, especially when it is done numerically, one can quickly perform a dimensionality check
\begin{align}
    \dim([n_{13},n_{23},0]) = \sum_L \kappa_{\max}(L) (2L+1),
\end{align}
since $M$ takes values $L,L-1,...,-L+1,-L$.

\vskip.2cm

From the expansion of the highest weight state in terms of Gelfand basis states, we can find the expansion of all other states using Eq. (\ref{eq:lower-weight})
\begin{align}
     \ket{\begin{array}{c}
n_{13},n_{23},0\cr
\kappa L M=L-\gamma
\end{array}} &= \sqrt{\frac{2^\gamma (2L-\gamma)!}{(2L)!\gamma!}}  \sum_{t=0}^{n_{23}} \sum_{q=q_{\min}(L,t)}^{q_{\max}(L,t)} C^{\kappa}_{qt}([n_{13},n_{23},0]LM=L) (E_{23}+E_{31})^{\gamma} \ket{\begin{array}{c}
n_{13},n_{23},0\cr
L+2q-t,t \cr
L+q
\end{array}} \nonumber \\
&= \sqrt{\frac{2^\gamma (2L-\gamma)!}{(2L)!\gamma!}} \sum_G C_{qt}^\kappa ([n_{13},n_{23},0]LM=L-\gamma)\ket{G},
\end{align}
where $\gamma=L-M=0,1,...,2L$ and $\ket{G}$'s are Gelfand states resulting from the action of $E_{23}$ and $E_{31}$ on the Gelfand states involved in the expansion of the highest weight state. Let us note that since it is possible to define the action of the U(3) generators, Eqs. (\ref{eq:remain}) and (\ref{eq:raise-lower}), as methods associated with a data structure that is used to define the Gelfand states in object-oriented programming languages like C++, we neither provide an explicit expression here for the expansion coefficients $C_{qt}^\kappa ([n_{13},n_{23},0]LM=L-\gamma)$ appearing in the above equation nor use it in our code. Nonetheless, should one be interested, explicit formulas for these coefficients are given by Eqs. (61) and (62) in the previous work \cite{Pan2016NPA}.

\subsection{SU(3)$\supset$SO(3)$\supset$SO(2) Wigner Coefficients}
A general Wigner coefficient $\rm SU(3)\supset SO(3) \supset SO(2)$ can be factorized into a reduced (double-barred) coefficient -- also called stretched coefficient -- multiplied with a usual $\rm SO(3)\supset SO(2)$ CGC:
\begin{multline}
    \braket{\begin{array}{c}
n_{13},n_{23},0\cr
\kappa L M
\end{array}; \begin{array}{c}
n_{13}',n_{23}',0\cr
\kappa'L'M'
\end{array}}{\begin{array}{c}
n_{13}''-n_{33}'',n_{23}''-n_{33}'',0\cr
\kappa''L''M''
\end{array}}_\rho = \braket{LM;L'M'}{L''M''} \times \\
\bra{\begin{array}{c}
n_{13},n_{23},0\cr
\kappa L 
\end{array}; \begin{array}{c}
n_{13}',n_{23}',0\cr
\kappa'L'
\end{array}} \text{} \ket{\begin{array}{c}
n_{13}''-n_{33}'',n_{23}''-n_{33}'',0\cr
\kappa''L''
\end{array}}_\rho.
\end{multline}
On the grounds that $\rm SO(3)\supset SO(2)$ CGC's are well known, and that there are many computer programs that provide their evaluation, we only concentrate on the reduced coefficients in this paper. 

\vskip.2cm

The formula for  the evaluation of $\rm SU(3)\supset SO(3)$ Wigner coefficients in terms of $\rm U(3) \supset U(2) \supset U(1)$ CGC's is provided in \cite{Pan2016NPA}; we employ that formula here too, however, without factorizing into $\rm U(3)\supset U(2)$ and $\rm U(2)\supset U(1)$, since the entire coefficients can be computed using the method described in the previous subsection:
\begin{multline}
\bra{\begin{array}{c}
n_{13},n_{23},0\cr
\kappa L 
\end{array}; \begin{array}{c}
n_{13}',n_{23}',0\cr
\kappa'L'
\end{array}} \text{} \ket{\begin{array}{c}
n_{13}''-n_{33}'',n_{23}''-n_{33}'',0\cr
\kappa''L''
\end{array}}_\rho = \braket{LL-\gamma;L'L'}{L''L''}^{-1} \sqrt{\frac{(2L-\gamma)!2^\gamma}{(2L)!\gamma!}} \times  \\
 \sum_{q_i,t_i} C_{q_1 t_1}^{\kappa}([n_{13},n_{23},0]LM=L-\gamma) C_{q_2 t_2}^{\kappa'}([n_{13}',n_{23}',0]L'M'=L') C_{q_3-n_{33}'', t_3-n_{33}''}^{\kappa''}([n_{13}''-n_{33}'',n_{23}''-n_{33}'',0]L''M''=L'') \\
\times \bra{\begin{array}{c}
n_{13},n_{23},0\cr
L-\gamma+2q_1-t_1,t_1 \cr
L-\gamma+q_1
\end{array}; \begin{array}{c}
n_{13}',n_{23}',0\cr
L'+2q_2-t_2,t_2\cr
L'+q_2
\end{array}}\ket{\begin{array}{c}
n_{13}''-n_{33}'',n_{23}''-n_{33}'',0\cr
L''+2q_3-t_3,t_3\cr
L''+q_3
\end{array}}_\rho,
\end{multline}
where $\gamma=L+L'-L''=0,1,...,L$.

\subsection{Summary of the Algorithm for Computing $\rm SU(3) \supset SO(3)$  Wigner Coefficients}
The procedure to compute all $\rm SU(3) \supset SO(3)$ Wigner coefficients associated with the tensor product of two SU(3) irreps, $[n_{13},n_{23}] \otimes [n_{13}',n_{23}']$ or in Elliott's notation $(\lambda,\mu) \otimes (\lambda',\mu')$, can be summarized as follows,
\begin{enumerate}
    \item Transform the two SU(3) irreps to two U(3) equivalent irreps $[n_{13},n_{23},0] \otimes [n_{13}',n_{23}',0]$ and then find all coupled irreps that occur in the decomposition of the tensor product, $[n_{13}'',n_{23}'',n_{33}'']$ as described in \cite{Alex2011JMP}.
    \item Follow the procedure in Subsection VI.B to find the expansion of the angular momentum highest weight states, $\ket{[n_{13},n_{23},0]\kappa LM=L},\ket{[n_{13}',n_{23}',0]\kappa' L'M'=L'}$ and $\ket{[n_{13}''-n_{33}'',n_{23}''-n_{33}'',0]\kappa'' L''M''=L''}$, in their corresponding canonical Gelfand bases; and then the expansion of lower weight states $\ket{[n_{13},n_{23},0]\kappa LM}$  for the first irrep .
    \item Compute all $\rm U(3)\supset U(2) \supset U(1)$ CGC's for the highest weight states (Subsection IV.C) and then for all lower weight states (Subsection IV.D).
    \item Combine the results of steps 2 and 3 to calculate all Wigner coefficients as shown in Subsection VI.C.
\end{enumerate}
Note that steps 2 and 3 are independent of each other and their order can be switched without changing any results. Moreover, we believe that this new algorithm is more advantageous than that of Draayer and Akiyama \cite{Draayer1973JMP,Akiyama1973} due to the fact that it does not involve alternating sums and binomial coefficients which may introduce more numerical errors and slow down computations.

\section{Conclusion}

In this paper, we first present a very simple procedure for the evaluation of $\rm U(3)\supset U(2) \supset U(1)$ Clebsch-Gordan coefficients, which is based on the null space concept of the U(3) generators presented first by Arne Alex et al. \cite{Alex2011JMP}. The highest weight CGC's are obtained simultaneously from the solution of the null space of the two generators $E_{12}$ and $E_{23}$, whereas the lower weight CGC's can be obtained recursively using the counterpart generators $E_{21}$ and $E_{32}$. The method to compute CGC's in this canonical scheme is purely a mathematical tool and hence applicable to any physical system that respects U(3) symmetry or follows all the properties of the U(3) algebra. The calculation of Racah recoupling coefficients, which are 6-U(3) and 9-U(3) coefficients, is also discussed.

\vskip.2cm

Second, we demonstrate the application of U(3) -- and SU(3) -- to nuclear structure studies, which comes with the reduction $\rm SU(3) \supset SO(3)$ and Wigner coefficients associated thereto . This new methodology greatly benefits from the previous work \cite{Pan2016NPA} for the transformation between the physical basis states and the canonical ones. Since numerical algorithms for solving the null spaces of both dense and sparse matrices have been well-developed and fine-tuned using modern programming languages, an early laptop version of a new C++ library for evaluating all Wigner $\rm SU(3) \supset SO(3)$ coefficients that resolves all multiplicity issues, and also provides dimension checking, is now available and has demonstrated promising performance. Moreover, we believe this procedure will be able to generate faster and more accurate results in comparison with the 1973 formulation of Draayer and Akiyama \cite{Draayer1973JMP,Akiyama1973}, because the method herein avoids nested alternating sums and several binomial coefficients of huge factorials that stem from the Hill-Wheeler integral and Wigner D-functions involved in the 1973 work.

\vskip .5cm

\begin{acknowledgments}
{Support from the National Natural Science Foundation of China (12175097) is acknowledged; and from LSU through its Sponsored Research Rebate Program as well as under the LSU Foundation's Distinguished Research Professorship Program. And in addition, the authors wish to acknowledge the support many graduate students, some domestic but most from abroad, and especially the ongoing support of LSU Professors Kristina Launey (originally from Bulgaria) and Alexis Mercenne (originally from France) in addition to two external colleagues, Professor Feng Pan (from China) who is a co-author on this paper as well as Tomas Dytrych (from the Czech Republic) who was and still is the Chief Architect of the so-called symmetry-adapted no-core shell model (SA-NCSM) which is currently considered to be a theory of choice for the carrying out many-particle shell-model calculations in nuclear physics.}
\end{acknowledgments}

\bibliographystyle{ieeetr}
\bibliography{extracted_bib}

\end{document}